\documentclass[11pt]{article}
\usepackage{amssymb,amsmath,amsfonts}
\usepackage{graphicx}
\usepackage{graphics}
\usepackage{eepic,epsfig}

 

\begin{document}

\title{Transition radiation on a dynamical periodical interface}
\author{A. R. Mkrtchyan$^{1,2}$, A. P. Potylitsyn$^{2,3}$, V. R. Kocharyan$%
^{1,2}$, A. A. Saharian$^{1}$ \\
%EndAName
\\
\textit{$^{1}$Institute of Applied Problems in Physics,}\\
\textit{25 Nersessian Street, 0014 Yerevan, Armenia}\vspace{0.3cm}\\
\textit{$^2$Tomsk Polytechnic University, 30 Lenin Ave., 634050 Tomsk, Russia%
}\vspace{0.3cm}\\
\textit{$^3$National Research Nuclear University MEPhI, 115409 Moscow, Russia%
}}
\maketitle

\begin{abstract}
We investigate the transition radiation on a periodically deformed interface
between two dielectric media. Under the assumption that the dielectric
permittivities of the media are close, a formula is derived for the
spectral-angular distribution of the radiated energy in the general case of
a nonstatic profile function for the separating boundary. In particular, the
latter includes the case of surface waves propagating along the boundary.
The numerical examples are given for triangular grating and for sinusoidal
profile. We show that instead of a single peak in the backward transition
radiation on a flat interface, for periodic interface one has a set of
peaks. The number and the locations of the peaks depend on the incidence
angle of the charge and on the period of the interface. The conditions are
specified for their appearance.
\end{abstract}

\bigskip

\section{Introduction}

The polarization of a medium by a moving charged particle gives rise to a
number of radiation processes. Well-known examples are Cherenkov, transition
and diffractions radiations. In particular, many aspects of transition
radiation, both theoretical and experimental, have been investigated in
numerous publications (for reviews see \cite{TerM72}). Transition radiation
is produced when a relativistic particle traverses an inhomogeneous medium.
Such radiation has a number of remarkable properties and at present it has
found many important applications. In particular, optical and extreme
ultraviolet transition radiation from metallic targets observed in backward
direction was used for the measurement of transverse size, divergence and
energy of electron and proton beams (see, for example, \cite{Lump09} and
references therein).

The modern accelerators allow us to produce electron beams consisting of
trains of short bunches with subpicosecond duration \cite{Mugg08, Sun10}.
The conventional diagnostic tools do not provide the required measurement
accuracy and the development of new reliable and economic diagnostic
techniques is the actual task. Such techniques based on different radiation
mechanisms (for instance, transition radiation) are widely used \cite{Schl00}%
. Reconstruction of a bunch profile is carried out using the Fourier
transformation of the measured spectrum of coherent transition radiation
\cite{Cast99}. Recently a technique is proposed based on the Smith-Purcell
radiation spectral measurements \cite{Dori02,Douc06}. The advantages of such
kind of diagnostics are due to the fact that the Smith-Purcell radiation
spectral distribution is quasimonochromatic. But because of the weak
radiation intensity the usage of the proposed technique meets with troubles
\cite{Blac09}. We propose to use the transition radiation from periodically
deformed targets instead of the Smith-Purcell radiation mechanism for beam
diagnostic. In this case the radiation intensity is much higher but, as we
show below, the spectral distribution becomes quasimonochromatic similar to
the Smith-Purcell radiation and the abovementioned advantages are retained.

The main part of the investigations of transition radiation considers flat
interfaces separating the media with different optical properties. Very few
works have been devoted to the transition radiation from rough surfaces \cite%
{Bagh76}. A periodic structure of the separating interface can serve as an
additional tool for the control of the spectral-angular distribution of the
transition radiation. In general, the problem does not allow an analytic
solution and approximate or numerical methods should be used. An integral
method to study transition radiation from interfaces with arbitrary profile
was considered in \cite{Taka00}. The method is a generalization of that
previously discussed in \cite{Berg73} for the case of the Smith-Purcell
radiation. Relatively simple and physically more transparent expressions for
the radiation intensity are obtained by using various approximations. In
particular, the investigation of the transition radiation from
regular-roughness interface is carried out in \cite{Bagh01} under the
assumption that the dielectric permittivities of the media separated by the
interface are close to each other. The same approximation scheme has been
used in \cite{Mkrt89} for the investigation of the radiation intensity from
a charge or bunches flying over a surface waves.

In the present paper we consider the transition radiation on an interface
with arbitrary periodic profile and for an arbitrary incidence angle of a
charged particle. In particular, the function describing the profile can
also be time dependent. This type of dynamical interface may be realized by
surface waves propagating on the boundary of two media. The investigation
will be done within the framework of the perturbation theory with respect to
the small difference of the dielectric permittivities of the media. In
particular, the corresponding condition is satisfied for the X-ray
transition radiation. Another example is that when for a part of the medium
the dielectric permittivity is changed by some external influences, for
example, intense electromagnetic waves or acoustic waves. These changes are
much smaller than the corresponding unperturbed permittivities and their
effects are well described by the perturbation theory. Transition radiation
on a superlattice created by acoustic waves has been discussed previously in
\cite{Grig98}. The above-mentioned perturbation theory in general form with
a number of applications is reviewed in \cite{Davi82}. The corresponding
results can in certain cases be used to estimate the effects in more
complicated configurations.

The paper is organized as follows. In the next section we obtain the formula
for the spectral-angular density of the radiated energy for a general
profile function. The formulas are given for separate polarizations, as well
as for the total intensity. Special cases, which include diffraction
gratings and surface waves excited on the interface, are discussed in
section \ref{sec:Special}. Numerical examples for the both angular and
spectral distributions of the radiated energy are considered in section \ref%
{sec:Numerical}. The main results are summarized in section \ref{sec:Conc}.

\section{Transition radiation for a general case of the profile function}

\label{sec:Gen}

First let us consider the general case of a medium with dielectric
permittivity%
\begin{equation}
\varepsilon =\varepsilon _{0}+\Delta \varepsilon (\mathbf{r},t),  \label{eps}
\end{equation}%
where $\varepsilon _{0}$ is a constant and $|\Delta \varepsilon (\mathbf{r}%
,t)|\ll |\varepsilon _{0}|$. Let us denote by $\mathbf{E}_{0}$ and $\mathbf{H%
}_{0}$ the electric and magnetic fields created by the radiation source in a
homogeneous medium with permittivity $\varepsilon _{0}$. Presenting the
corresponding fields in the medium with permittivity (\ref{eps}) in the form
$\mathbf{E}=\mathbf{E}_{0}+\Delta \mathbf{E}$ and $\mathbf{H}=\mathbf{H}%
_{0}+\Delta \mathbf{H}$ and substituting into the Maxwell equations, we can
see that, in the first approximation with respect to the small quantity $%
|\Delta \varepsilon (\mathbf{r},t)/\varepsilon _{0}|$, the fields $\Delta
\mathbf{E}$ and $\Delta \mathbf{H}$ obey the inhomogeneous Maxwell equations
with the charge density $\rho =\mathrm{div}(\Delta \varepsilon \mathbf{E}%
_{0})/(4\pi )$ and current density $j=\partial _{t}(\Delta \varepsilon
\mathbf{E}_{0})/(4\pi )$. Consequently, in the same approximation, the
spectral-angular distribution of the radiated energy is given by the formula
\cite{Davi82,Davi81}%
\begin{equation}
W_{\omega }d\omega d\Omega =\frac{(2\pi )^{4}\omega ^{4}\sqrt{\varepsilon
_{0}}}{4c^{3}}\sum_{l=1}^{2}|\mathbf{e}^{l}(\Delta \varepsilon \mathbf{E}%
_{0})_{\mathbf{k},\omega }|^{2}d\omega d\Omega ,  \label{Int}
\end{equation}%
where $\mathbf{e}^{l}$, $l=1,2$, are independent unit vectors of the
polarization, $\omega $ and $\mathbf{k}$ are the frequency and wave vector
of the radiated photon, $k=|\mathbf{k}|=\omega \sqrt{\varepsilon }/c$, $%
d\Omega $ is the solid angle element. In (\ref{Int}),%
\begin{equation}
(\Delta \varepsilon \mathbf{E}_{0})_{\mathbf{k},\omega }=\int \frac{dtd%
\mathbf{r}}{(2\pi )^{4}}\,e^{-i(\mathbf{k}\,\mathbf{r}-\omega t)}\Delta
\varepsilon (\mathbf{r},t)\mathbf{E}_{0}(\mathbf{r},t),  \label{Four1}
\end{equation}%
and the separate terms in the sum over $l$ give the radiated energy for the
corresponding polarization.

The Fourier transform appearing in (\ref{Int}) is expressed in terms of the
Fourier transforms of the separate factors as%
\begin{equation}
(\Delta \varepsilon \mathbf{E}_{0})_{\mathbf{k},\omega }=\int d\omega
^{\prime }d\mathbf{k}^{\prime }\,\Delta \varepsilon \left( \mathbf{k}-%
\mathbf{k}^{\prime },\omega -\omega ^{\prime }\right) \mathbf{E}_{0}\left(
\mathbf{k}^{\prime },\omega ^{\prime }\right) ,  \label{FT1}
\end{equation}%
where%
\begin{equation}
\Delta \varepsilon \left( \mathbf{k},\omega \right) =\int \frac{dtd\mathbf{r}%
}{(2\pi )^{4}}\,e^{-i(\mathbf{k}\,\mathbf{r}-\omega t)}\Delta \varepsilon (%
\mathbf{r},t).  \label{FTeps}
\end{equation}%
For a point charge $q$ moving with a constant velocity $\mathbf{v}$ along
the trajectory given by $\mathbf{r}=\mathbf{r}_{0}+\mathbf{v}(t-t_{0})$, the
Fourier transform of the corresponding electric field is determined by the
expression%
\begin{equation}
\mathbf{E}_{0}(\mathbf{k},\omega )=\frac{4\pi iq}{(2\pi )^{3}}\frac{\omega
\mathbf{v}/c^{2}-\mathbf{k}/\varepsilon _{0}}{k^{2}-\omega ^{2}\varepsilon
_{0}/c^{2}}e^{-i(\mathbf{k}\,\mathbf{r}_{0}-\omega t_{0})}\delta (\omega -%
\mathbf{k}\,\mathbf{v}).  \label{FTE0}
\end{equation}

In the present paper we consider the transition radiation on the boundary of
two homogeneous media with dielectric permittivities $\varepsilon _{0}$ and $%
\varepsilon _{1}$. For a general case of dynamical boundary the
corresponding equation can be written as
\begin{equation}
x=x_{0}(y,z,t).  \label{x0}
\end{equation}%
Here and below $(x,y,z)$ stand for the Cartesian coordinates. Hence, in the
problem under consideration the dielectric permittivity is given by%
\begin{equation}
\varepsilon =\left\{
\begin{array}{ll}
\varepsilon _{0}, & x<x_{0}(y,z,t), \\
\varepsilon _{1}, & x>x_{0}(y,z,t).%
\end{array}%
\right.  \label{eps2}
\end{equation}%
For the function $\Delta \varepsilon (\mathbf{r},t)$ in (\ref{eps}) one can
take%
\begin{equation}
\Delta \varepsilon (\mathbf{r},t)=\Delta \varepsilon \,\theta
(x-x_{0}(y,z,t)),\;\Delta \varepsilon =\varepsilon _{1}-\varepsilon _{0},
\label{Deleps}
\end{equation}%
where $\theta (x)$ is the Heaviside unit step function. Assuming that $%
|\Delta \varepsilon |/\varepsilon _{0}\ll 1$, for the Fourier transform (\ref%
{FT1}), under the condition $\omega \neq \mathbf{kv}$, one gets%
\begin{eqnarray}
(\Delta \varepsilon \mathbf{E}_{0})_{\mathbf{k},\omega } &=&\frac{4\pi q}{%
(2\pi )^{7}}\int d\mathbf{k}^{\prime }\,\frac{\Delta \varepsilon }{K_{x}}%
\frac{\left( \mathbf{k}^{\prime }\mathbf{v}\right) \mathbf{v}/c^{2}-\mathbf{k%
}^{\prime }/\varepsilon _{0}}{k^{\prime 2}-\left( \mathbf{k}^{\prime }%
\mathbf{v}\right) ^{2}\varepsilon _{0}/c^{2}}e^{-i\mathbf{k}^{\prime }(\,%
\mathbf{r}_{0}-\mathbf{v}t_{0})}  \notag \\
&&\times \int dtdydz\,e^{i((\omega -\mathbf{k}\,^{\prime }\mathbf{v}%
)t+K_{y}y+K_{z}z)}e^{iK_{x}x_{0}(y,z,t)},  \label{Four2}
\end{eqnarray}%
where%
\begin{equation}
\mathbf{K}=\mathbf{k}^{\prime }-\mathbf{k}.  \label{Kbf}
\end{equation}%
For $\omega =\mathbf{kv}$ an additional term is present in the expression
for $(\Delta \varepsilon \mathbf{E}_{0})_{\mathbf{k},\omega }$ containing
the factor $\delta (\omega -\mathbf{kv})$. This term gives contribution to
the radiation along the Cherenkov angle only, if the corresponding condition
is satisfied.

Assuming that the function $x_{0}(y,z,t)$ is periodic with the periods $%
L_{y} $, $L_{z}$, and $T$, for the function in the integrand of (\ref{Four2}%
) we can write the Fourier expansion%
\begin{equation}
e^{iK_{x}\,x_{0}(y,z,t)}=\sum_{m,n,l=-\infty }^{+\infty
}f_{mnl}(K_{x})e^{2\pi i\left( my/L_{y}+nz/L_{z}\right) -i\omega _{l}t},
\label{Four5}
\end{equation}%
where $\omega _{l}=2\pi l/T$, and%
\begin{equation}
f_{mnl}(K_{x})=\frac{1}{L_{y}L_{z}T}\int_{0}^{L_{y}}dy\int_{0}^{L_{z}}dz%
\int_{0}^{T}dt\,e^{iK_{x}\,x_{0}(y,z,t)}e^{-2\pi i\left(
my/L_{y}+nz/L_{z}\right) +i\omega _{l}t}.  \label{FourIm}
\end{equation}%
Substituting the expansion (\ref{Four5}) into (\ref{Four2}) one gets%
\begin{equation}
(\Delta \varepsilon \mathbf{E}_{0})_{\mathbf{k},\omega }=\frac{2q\Delta
\varepsilon }{(2\pi )^{3}v_{x}\varepsilon _{0}}\,\sum_{m,n,l=-\infty
}^{+\infty }\frac{f_{mnl}(K_{x})}{K_{x}}\frac{\mathbf{k}^{\prime }-\left(
\omega \mathbf{-}\omega _{l}\right) \mathbf{v}\varepsilon _{0}/c^{2}}{%
k^{\prime 2}-\left( \omega \mathbf{-}\omega _{l}\right) ^{2}\varepsilon
_{0}/c^{2}}e^{-i\mathbf{k}^{\prime }(\,\mathbf{r}_{0}-\mathbf{v}t_{0})},
\label{Four6}
\end{equation}%
with the notations%
\begin{eqnarray}
\mathbf{k}^{\prime } &=&\mathbf{k}+K_{x}\mathbf{e}_{x}-\mathbf{g},  \notag \\
K_{x} &=&\frac{1}{v_{x}}\left( \omega -\mathbf{k}\,\mathbf{v}-\omega _{l}+%
\mathbf{g}\,\mathbf{v}\right) ,  \notag \\
\mathbf{g} &=&(0,2\pi m/L_{y},2\pi n/L_{z}),  \label{Kx}
\end{eqnarray}%
where $\mathbf{e}_{x}$ is the unit vector along the $x$-axis. Note that we
have the relation%
\begin{equation}
\mathbf{k}^{\prime }\mathbf{v=}\omega \mathbf{-}\omega _{l}.  \label{kv}
\end{equation}

Let us denote the angle between the vectors $\mathbf{v}$ and $\mathbf{k}$ by
$\theta _{0}$. Then, for the polarization vectors of the parallel and
perpendicular polarizations one has%
\begin{equation}
\mathbf{e}^{1}=\mathbf{e}^{\parallel }=\frac{\mathbf{v}-\left( \mathbf{n}\,%
\mathbf{v}\right) \mathbf{n}}{v\sin \theta _{0}},\;\mathbf{e}^{2}=\mathbf{e}%
^{\perp }=\frac{\mathbf{n}\times \mathbf{v}}{v\sin \theta _{0}},
\label{PolVec}
\end{equation}%
where $\mathbf{n}=\mathbf{k}/k$ is the unit vector along the radiation
direction. For the corresponding spectral-angular densities of the radiated
energy one has%
\begin{equation}
W_{\omega }^{p}=\frac{(2\pi )^{4}\omega ^{4}\sqrt{\varepsilon _{0}}}{4c^{3}}|%
\mathbf{e}^{p}(\Delta \varepsilon \mathbf{E}_{0})_{\mathbf{k},\omega }|^{2},
\label{Wp}
\end{equation}%
with $p=\parallel $ and $p=\perp $. The expression for the total
spectral-angular density takes the form%
\begin{equation}
W_{\omega }=W_{\omega }^{\parallel }+W_{\omega }^{\perp }=\frac{(2\pi
)^{4}\omega ^{4}\sqrt{\varepsilon _{0}}}{4c^{3}}\left[ \left\vert (\Delta
\varepsilon \mathbf{E}_{0})_{\mathbf{k},\omega }\right\vert ^{2}-\left\vert (%
\mathbf{n}(\Delta \varepsilon \mathbf{E}_{0})_{\mathbf{k},\omega
})\right\vert ^{2}\right] .  \label{W}
\end{equation}%
Note that we can also write%
\begin{equation}
\left\vert (\Delta \varepsilon \mathbf{E}_{0})_{\mathbf{k},\omega
}\right\vert ^{2}-\left\vert (\mathbf{n}(\Delta \varepsilon \mathbf{E}_{0})_{%
\mathbf{k},\omega })\right\vert ^{2}=\left\vert \mathbf{n}\times (\Delta
\varepsilon \mathbf{E}_{0})_{\mathbf{k},\omega }\right\vert ^{2}.  \label{nE}
\end{equation}

Substituting the expression (\ref{Four6}) into (\ref{Wp}) and assuming that
the beam cross section is larger than the periods $L_{y}$ and $L_{z}$, for
the spectral-angular densities of the radiated energy on separate
polarizations, averaged over the impact parameter, one gets the expression%
\begin{equation}
W_{\omega }^{p}=\frac{q^{2}\left( \Delta \varepsilon \right) ^{2}\omega ^{4}%
}{4\pi ^{2}c^{3}\varepsilon _{0}^{3/2}v_{x}^{2}v^{2}\sin ^{2}\theta _{0}}%
\,\sum_{m,n,l=-\infty }^{+\infty }\frac{V_{p}|f_{mnl}(K_{x})|^{2}/K_{x}^{2}}{%
\left[ \omega _{l}(2\omega -\omega _{l})\varepsilon _{0}/c^{2}+K_{x}^{2}+%
\mathbf{g}^{2}+2\left( k_{x}K_{x}-\mathbf{kg}\right) \right] ^{2}},
\label{Wp3}
\end{equation}%
where for the parallel and perpendicular polarizations we have%
\begin{eqnarray*}
V_{\parallel } &=&\left\{ \left( \omega \mathbf{-}\omega _{l}\right) \left(
1-\beta _{0}^{2}\right) -\left( \mathbf{n}\,\mathbf{v}\right) [\omega \sqrt{%
\varepsilon _{0}}/c+n_{x}K_{x}-\mathbf{ng}-\left( \omega \mathbf{-}\omega
_{l}\right) \left( \mathbf{n}\,\mathbf{v}\right) \varepsilon
_{0}/c^{2}]\right\} ^{2}, \\
V_{\perp } &=&\left( [\mathbf{n}\times \mathbf{v}]\cdot \left( K_{x}\mathbf{e%
}_{x}-\mathbf{g}\right) \right) ^{2},
\end{eqnarray*}%
with the notation $\beta _{0}=v\sqrt{\varepsilon _{0}}/c$. The
spectral-angular density for the total radiation is given by the formula%
\begin{equation}
W_{\omega }=\frac{q^{2}\left( \Delta \varepsilon \right) ^{2}\omega ^{4}}{%
(2\pi )^{2}c^{3}v_{x}^{2}\varepsilon _{0}^{3/2}}\sum_{m,n,l=-\infty
}^{+\infty }\frac{V|f_{mnl}(K_{x})|^{2}/K_{x}^{2}}{\left[ \omega
_{l}(2\omega -\omega _{l})\varepsilon _{0}/c^{2}+K_{x}^{2}+\mathbf{g}%
^{2}+2\left( k_{x}K_{x}-\mathbf{kg}\right) \right] ^{2}},  \label{W2}
\end{equation}%
with the function%
\begin{eqnarray}
V &=&K_{x}^{2}+\mathbf{g}^{2}-\left[ (2-\beta _{0}^{2})\left( \omega \mathbf{%
-}\omega _{l}\right) -2\left( \mathbf{k}\,\mathbf{v}\right) \right] \left(
\omega \mathbf{-}\omega _{l}\right) \varepsilon _{0}/c^{2}  \notag \\
&&-\left[ K_{x}n_{x}-\mathbf{ng}-\left( \omega \mathbf{-}\omega _{l}\right)
\mathbf{nv}\varepsilon _{0}/c^{2}\right] ^{2}.  \label{V}
\end{eqnarray}%
The presented formulas are valid in the first approximation with respect to
the ratio $\Delta \varepsilon /\varepsilon _{0}$, for both backward and
inward radiations. An additional condition, that is obtained comparing with
the exact expressions in the case of flat boundary, is given below in
section \ref{sec:Special}.

Special cases of the profile function for the interface and numerical
examples will be discussed below. However, some features can be seen from
general formulas. The spectral-angular density of the radiated energy
contains the square of the vector (see (\ref{Four6}))%
\begin{equation}
\mathbf{A}(\mathbf{k}^{\prime })=\frac{\mathbf{k}^{\prime }-\left( \omega
\mathbf{-}\omega _{l}\right) \mathbf{v}\varepsilon _{0}/c^{2}}{k^{\prime
2}-\left( \omega \mathbf{-}\omega _{l}\right) ^{2}\varepsilon _{0}/c^{2}}.
\label{Ak}
\end{equation}%
Introducing the angle $\theta ^{\prime }$ between the vectors $\mathbf{k}%
^{\prime }$ and $\mathbf{v}$ and by taking into account the relation (\ref%
{kv}), we can see that%
\begin{equation}
|\mathbf{A}(\mathbf{k}^{\prime })|^{2}=\frac{1-\left( 2-\beta
_{0}^{2}\right) \beta _{0}^{2}\cos ^{2}\theta ^{\prime }}{k^{\prime 2}\left(
1-\beta _{0}^{2}\cos ^{2}\theta ^{\prime }\right) ^{2}}.  \label{Ak2}
\end{equation}%
From here it follows that if $\beta _{0}^{2}$ is close to 1, $|1-\beta
_{0}^{2}|\ll 1$, peaks can appear in the spectral angular distribution of
the radiated energy for small values of the angle $\theta ^{\prime }$. At
these peaks one has%
\begin{equation}
|\mathbf{A}(\mathbf{k}^{\prime })|^{2}\approx \frac{\theta ^{\prime 2}\gamma
_{0}^{4}}{k^{\prime 2}\left( 1+\gamma _{0}^{2}\theta ^{\prime 2}\right) ^{2}}%
,  \label{Ak2peak}
\end{equation}%
where $\gamma _{0}^{2}=1/(1-\beta _{0}^{2})$, $|\gamma _{0}^{2}|\gg 1$. For $%
\varepsilon _{0}=1$ the parameter $\gamma _{0}$ coincides with the gamma
factor of the radiating particle. One has $\theta ^{\prime }\sim 1/\gamma
_{0}$ and the spectral-angular density at the peaks increases with
increasing energy as $\gamma _{0}^{2}$. The integration over the angle will
give an additional factor $\theta ^{\prime }$ and the spectral density at
the peaks increases as $\gamma _{0}$.

The condition for the angle $\theta ^{\prime }$ to be small can be
translated in terms of the wave vector of the radiated photon as%
\begin{equation}
(\mathbf{k}-\mathbf{g}-\frac{\omega \mathbf{-}\omega _{l}}{v^{2}}\mathbf{v}%
)\times \mathbf{e}_{x}=\mathcal{O}(1/\gamma _{0}^{2}).  \label{CondPeak}
\end{equation}%
This condition determines the location of possible peaks in the
spectral-angular density of the radiated energy up to the terms of the order
$1/\gamma _{0}^{2}$. Examples of this type of peaks will be given below.
Note that the condition (\ref{CondPeak}) does not depend on the form of the
profile function and is completely determined by the periodicity properties
of the interface.

In deriving the expressions for the spectral-angular distribution of the
radiation intensity we have assumed that the charge of the radiating
particle is fixed. New features may arise in the case of the transition
radiation from multiply charged ions. In particular, as a result of the
interaction with medium the ion can gain or lose charge. The transition
radiation on a flat interface, by taking into account this effect, has been
recently discussed in \cite{Maly15}. It has been shown that the change in
the charge of the particle can lead to a considerable increase of the
radiation intensity.

\section{Special cases}

\label{sec:Special}

As a check of the general formulas given above, first let as consider the
case of a flat boundary between the media. In this case one has $%
x_{0}(y,z,t)=x_{0}=\mathrm{const}$ and the $x$-axis is perpendicular to the
boundary. By taking into account that $f_{mnl}(K_{x})=\delta _{m0}\delta
_{n0}\delta _{l0}e^{iK_{x}\,x_{0}}$ and introducing the angles $\theta _{v}$
and $\theta $ in accordance with $v_{x}=v\cos \theta _{v}$ and $k_{x}=k\cos
\theta $, for the spectral angular density of the radiated energy we get%
\begin{equation}
W_{\omega }^{(\mathrm{flat})}=\frac{q^{2}\left( \Delta \varepsilon \right)
^{2}v^{2}\cos ^{2}\theta _{v}}{(2\pi )^{2}c^{3}\varepsilon _{0}^{3/2}}\frac{%
w^{2}-\left( 2w-\beta _{0}^{2}\right) \beta _{0}^{2}\cos ^{2}\theta
_{v}-\left( w\cos \theta -\beta _{0}^{2}\cos \theta _{v}\cos \theta
_{0}\right) ^{2}}{w^{4}\left( w+2\beta _{0}\cos \theta _{v}\cos \theta
\right) ^{2}},  \label{Wflat}
\end{equation}%
where%
\begin{equation}
w=1-\beta _{0}\cos \theta _{0}.  \label{w}
\end{equation}%
The special case of normal incidence is obtained taking $\theta _{v}=\pi $, $%
\theta =\pi -\theta _{0}$. It can be seen that (\ref{Wflat}) coincides with
the exact formula in the limit when $\Delta \varepsilon $ is small and,
additionally, under the conditions
\begin{equation}
|\Delta \varepsilon /\varepsilon _{0}|\ll \cos ^{2}\theta ,\;|\Delta
\varepsilon /\varepsilon _{0}|\ll |\left( 1-\beta _{0}\cos \theta
_{0}\right) \cos \theta /\cos \theta _{v}|.  \label{CondAd}
\end{equation}%
In particular, from these conditions it follows that the expressions given
above are not valid for the radiation propagating nearly parallel to the
interface (the angle $\theta $ is close to $\pi /2$). Limiting formulas,
similar to (\ref{Wflat}), can also be obtained for the densities of separate
polarizations.

Note that for the $l=m=n=0$ term in the general formula (\ref{W2}) we have%
\begin{equation}
W_{\omega }(l=m=n=0)=|f_{000}(K_{x})|^{2}W_{\omega }^{(\mathrm{flat})}.
\label{Wlmn0}
\end{equation}%
By taking into account that $|f_{000}(K_{x})|^{2}\leqslant 1$, we conclude
that the corresponding contribution to the radiation cannot exceed $%
W_{\omega }^{(\mathrm{flat})}$. In the case of a flat interface and in the
limit $|1-\beta _{0}|\ll 1$, the peaks in the spectral-angular distribution
of the radiation are realized for small angles $\theta _{0}$ and near the
angle for which the factor $\left( w+2\beta _{0}\cos \theta _{v}\cos \theta
\right) ^{2}$ in the denominator of (\ref{Wflat}) takes its minimal value.
The first one corresponds to the peak in the forward radiation located near
the particle velocity and the second one corresponds to the direction of
specular reflection ($\theta =\pi -\theta _{v}$, $\theta _{0}=\pi -2\theta $%
).

Now we turn to the case of the profile function
\begin{equation}
x_{0}(y,z,t)=x_{0}(z-v_{s}t),  \label{x0Run}
\end{equation}%
where the function $x_{0}(u)$ is periodic with the period $L$. Note that the
function (\ref{x0Run}) describes a running surface wave with the wavelength $%
L$ and velocity $v_{s}=L/T$. In this case for the corresponding function $%
f_{mnl}(K_{x})$ one has%
\begin{equation}
f_{mnl}(K_{x})=\delta _{m0}\delta _{nl}f_{l}(K_{x}),  \label{fmnlSp}
\end{equation}%
with%
\begin{equation}
f_{l}(K_{x})=\frac{1}{L}\int_{0}^{L}dz\,e^{iK_{x}\,x_{0}(z)-2\pi ilz/L}.
\label{fnKx}
\end{equation}%
The corresponding radiation intensities are given by the expressions (\ref%
{Wp3}) and (\ref{W2}) with the replacements%
\begin{equation}
\sum_{m,n,l=-\infty }^{+\infty }\rightarrow \sum_{l=-\infty }^{+\infty
},\;f_{mnl}(K_{x})\rightarrow f_{l}(K_{x}),  \label{ReplSp}
\end{equation}%
and with%
\begin{equation}
\mathbf{g}=g_{l}\mathbf{e}_{z}=\left( 0,0,g_{l}\right) ,\;g_{l}=2\pi l/L,
\label{gSp}
\end{equation}%
where $\mathbf{e}_{z}$ is the unit vector along the $z$-axis.

Introducing spherical angular coordinates for the vectors $\mathbf{k}$ and $%
\mathbf{v}$,%
\begin{eqnarray}
\mathbf{k} &=&\frac{\omega }{c}\sqrt{\varepsilon _{0}}\left( \cos \theta
,\sin \theta \sin \phi ,\sin \theta \cos \phi \right) ,  \notag \\
\mathbf{v} &=&v\left( \cos \theta _{v},\sin \theta _{v}\sin \phi _{v},\sin
\theta _{v}\cos \phi _{v}\right) ,  \label{kvang}
\end{eqnarray}%
the spectral-angular density of the radiated energy on separate
polarizations, for the profile function (\ref{x0Run}), are presented as%
\begin{equation}
W_{\omega }^{p}=\frac{q^{2}\left( \Delta \varepsilon \right)
^{2}v^{2}\varepsilon _{0}^{-3/2}}{4\pi ^{2}c^{3}\cos ^{2}\theta _{v}\sin
^{2}\theta _{0}}\,\sum_{l=-\infty }^{+\infty }\frac{U_{p}^{2}|f_{l}(\omega
K/v)|^{2}}{K^{2}U^{2}},  \label{Wpang}
\end{equation}%
with the notations%
\begin{equation}
K=\frac{1}{\cos \theta _{v}}\left( 1-\beta _{0}\cos \theta _{0}-\frac{\omega
_{l}}{\omega }+\frac{vg_{l}}{\omega }\sin \theta _{v}\cos \phi _{v}\right) ,
\label{K}
\end{equation}%
and%
\begin{eqnarray}
U &=&\frac{\omega _{l}}{\omega }\left( 2-\frac{\omega _{l}}{\omega }\right)
\beta _{0}^{2}+K^{2}+\frac{v^{2}}{\omega ^{2}}g_{l}^{2}+2\beta _{0}\left(
K\cos \theta -\frac{vg_{l}}{\omega }\sin \theta \cos \phi \right) ,  \notag
\\
U_{\parallel } &=&\left( 1\mathbf{-}\frac{\omega _{l}}{\omega }\right)
\left( 1-\beta _{0}^{2}\right) -\cos \theta _{0}\left[ \beta _{0}+K\cos
\theta -\frac{vg_{l}}{\omega }\sin \theta \cos \phi -\beta _{0}^{2}\left( 1%
\mathbf{-}\frac{\omega _{l}}{\omega }\right) \cos \theta _{0}\right] ,
\notag \\
U_{\perp } &=&[\mathbf{n}\times \mathbf{v}/v]\cdot \left( K\mathbf{e}%
_{x}-(vg_{l}/\omega )\mathbf{e}_{z}\right) .  \label{U}
\end{eqnarray}%
For the total density we get%
\begin{equation}
W_{\omega }=\frac{q^{2}\left( \Delta \varepsilon \right)
^{2}v^{2}\varepsilon _{0}^{-3/2}}{4\pi ^{2}c^{3}\cos ^{2}\theta _{v}}%
\sum_{l=-\infty }^{+\infty }\frac{U_{t}|f_{l}(\omega K/v)|^{2}}{K^{2}U^{2}},
\label{Wang}
\end{equation}%
where%
\begin{eqnarray}
U_{t} &=&K^{2}+\frac{v^{2}g_{l}^{2}}{\omega ^{2}}-\left[ (2-\beta
_{0}^{2})\left( 1\mathbf{-}\frac{\omega _{l}}{\omega }\right) -2\beta
_{0}\cos \theta _{0}\right] \left( 1\mathbf{-}\frac{\omega _{l}}{\omega }%
\right) \beta _{0}^{2}  \notag \\
&&-\left[ K\cos \theta -\frac{vg_{l}}{\omega }\sin \theta \cos \phi -\left( 1%
\mathbf{-}\frac{\omega _{l}}{\omega }\right) \beta _{0}^{2}\cos \theta _{0}%
\right] ^{2}.  \label{Ut}
\end{eqnarray}%
Note that $vg_{l}/\omega =\beta _{0}l\lambda /L$, where $\lambda $ is the
radiation wavelength. In the case of a surface wave one has $\omega
_{l}=l\omega _{s}$, with $\omega _{s}=2\pi /T$ being the frequency of the
surface wave. For the radiation on frequencies $\omega \gg \omega _{s}$, the
terms containing the ratio $\omega _{l}/\omega $ can be omitted and the
spectral-angular densities coincide with the ones for the radiation on a
static interface with the same profile function.

Let us denote by $a$ the amplitude of the profile function $x_{0}(z)$. We
can estimate the dependence of the radiation intensity on $a$, for large
values of the ratio $a/\lambda $, by applying the stationary phase method to
the integral (\ref{fnKx}). If the function $x_{0}(z)$ has an stationary
point at $z=z_{0}$, $x_{0}^{\prime }(z_{0})=0$, then for $a/\lambda \gg 1$
the dominant contribution comes from the region of the integration near that
point. In this case one has $f_{l}(\omega K/v)\varpropto \sqrt{\lambda /a}$
and for large amplitudes the spectral-angular density decays as $1/a$. If
the function $x_{0}(z)$ has no stationary point then $f_{l}(\omega
K/v)\varpropto \lambda /a$ and the spectral-angular density behaves as $%
1/a^{2}$. For small values of the amplitude, $a/\lambda \ll 1$, the dominant
contribution to the radiation intensity comes from the $l=0$ term. The
leading term coincides with the corresponding quantity for a flat boundary.
The contributions from the terms $l\neq 0$ are suppressed by the factor $%
(a/\lambda )^{2}$.

\begin{figure}[tbph]
\begin{center}
\epsfig{figure=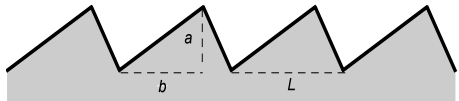,width=6.cm,height=1.7cm}
\end{center}
\caption{Parameters of the triangular grating.}
\label{fig1}
\end{figure}

The special case of the profile function (\ref{x0Run}) with $v_{s}=0$ ($%
T\rightarrow \infty $ for a fixed $L$) corresponds to a static periodic
interface with the period $L$. In this case one has $\omega _{l}=0$. As an
example of the separating boundary let us consider a triangular grating with
the parameters displayed in figure \ref{fig1}. The corresponding profile
function has the form%
\begin{equation}
x_{0}(z)=\left\{
\begin{array}{ll}
\frac{a}{b}(z-z_{0}), & z_{0}\leqslant z\leqslant z_{0}+b \\
\frac{a}{L-b}(z_{0}+L-z), & z_{0}+b\leqslant z\leqslant z_{0}+L.%
\end{array}%
\right.  \label{x0Grat}
\end{equation}%
With this function, for the Fourier transform (\ref{fnKx}) we obtain the
following expression%
\begin{equation}
f_{l}(K_{x})=2e^{i\,K_{x}a/2-\pi il\left( 2z_{0}+b\right) /L}\frac{%
K_{x}\,a\sin \left( K_{x}\,a/2-g_{l}b/2\right) }{\left(
K_{x}\,a-g_{l}b\right) \left( K_{x}\,a-g_{l}b+2\pi l\right) }.
\label{fnGrat}
\end{equation}%
For large values of the amplitude $a$, the function $|f_{l}(K_{x})|^{2}$
decays as $1/a^{2}$. This agrees with the general estimate given above, by
taking into account that the function (\ref{x0Grat}) has no stationary point.

As a next example of the profile function we consider a sinusoidal surface
wave%
\begin{equation}
x_{0}(z-v_{s}t)=a\sin (k_{s}z-\omega _{s}t),  \label{x0wave}
\end{equation}%
where $v_{s}=\omega _{s}/k_{s}$, $k_{s}=2\pi /L$. For this function, the
integral in (\ref{fnKx}) is expressed in terms of the Bessel function $%
J_{l}(x)$:
\begin{equation}
f_{l}(K_{x})=J_{l}(aK_{x})\,.  \label{flSurfWave}
\end{equation}%
For large values of the amplitude one has $|f_{l}(K_{x})|^{2}\varpropto 1/a$
which, again, is in agreement with the estimate given before for the general
case.

\section{Numerical analysis of the spectral-angular density}

\label{sec:Numerical}

\begin{figure}[tbph]
\begin{center}
\epsfig{figure=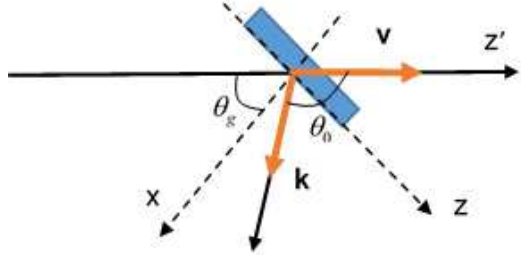,width=7.cm,height=3.5cm}
\end{center}
\caption{The geometry of the problem.}
\label{fig2}
\end{figure}

For the numerical analysis of the radiation intensity we will study the
relatively simple case when the vectors $\mathbf{v}$, $\mathbf{e}_{x}$ and $%
\mathbf{e}_{z}$ lie in the same plane and will consider the radiation
propagating in that plane (the vector $\mathbf{k}$ is in the plane formed by
$\mathbf{v}$ and $\mathbf{e}_{x}$, corresponding to $\phi =0$ and $\phi =\pi
$). The geometry of the problem is depicted in figure \ref{fig2}.
In this special case $W_{\omega }^{\perp }=0$ and the radiation is linearly
polarized: $W_{\omega }=W_{\omega }^{\parallel }$. The corresponding
expression is given by (\ref{Wang}) with $\phi _{v}=0$ and the expression
for $U$ is simplified to%
\begin{equation}
U=\frac{1}{\cos ^{2}\theta _{v}}\left[ \frac{vg_{l}}{\omega }-\beta _{0}\sin
\theta +\left( 1-\frac{\omega _{l}}{\omega }\right) \sin \theta _{v}\right]
^{2}+\left( 1-\frac{\omega _{l}}{\omega }\right) ^{2}\left( 1-\beta
_{0}^{2}\right) .  \label{U2}
\end{equation}%
Note that one has $\theta _{v}=\pi -\theta _{g}$, where $\theta _{g}$ is the
charge incidence angle with respect to the normal to the boundary ($x$%
-axis). In addition, we have $\theta =\pi -\theta _{g}-\theta _{0}$, $\phi =0
$ for $\theta _{0}<\pi -\theta _{g}$, and $\theta =\theta _{0}+\theta
_{g}-\pi $, $\phi =\pi $ for $\theta _{0}>\pi -\theta _{g}$. In the special
case under consideration, the condition (\ref{CondPeak}) for the appearance
of possible peaks is written as%
\begin{equation}
\beta _{0}\sin \left( \theta _{g}+\theta _{0}\right) -\sin \theta
_{g}=l\beta _{0}\frac{\lambda }{L}\left( 1-\frac{v_{s}}{v}\sin \theta
_{g}\right) +\mathcal{O}(1/\gamma _{0}^{2}).  \label{CondPeak2}
\end{equation}%
By taking into account that $vg_{l}/\omega =l\beta _{0}\lambda /L$, we see
that under this condition the expression in the square brackets of (\ref{U2}%
) is of the order $1/\gamma _{0}^{2}$ and, hence, at the peaks we have $%
U\sim 1/\gamma _{0}^{2}$. For a static profile one has $v_{s}=0$ and (\ref%
{CondPeak2}) reduces to the condition obtained in \cite{Poty00} for the
resonant diffraction radiation from a particle moving close to tilted
grating. For $\theta _{g}=\pi /2$ and $v_{s}=0$ one gets the standard
condition for the peaks of the Smith-Purcell radiation. From (\ref{CondPeak2}%
) it follows that, for a fixed value of the incidence angle $\theta _{g}$,
the number of possible values of $l$ for which the peaks appear increases
with increasing $L/\lambda $. Hence, instead of a single peak in the
backward transition radiation on a flat interface (in the direction of
specular reflection), in the case of periodic interface one has a set of
peaks. The number and the locations of the peaks depend on the incidence
angle of the charge and on the periodicity properties of the interface. In
the limit when the particle trajectory is parallel to the interface without
crossing we get the resonance peaks of the Smith-Purcell radiation.

\begin{figure}[tbph]
\begin{center}
\begin{tabular}{cc}
\epsfig{figure=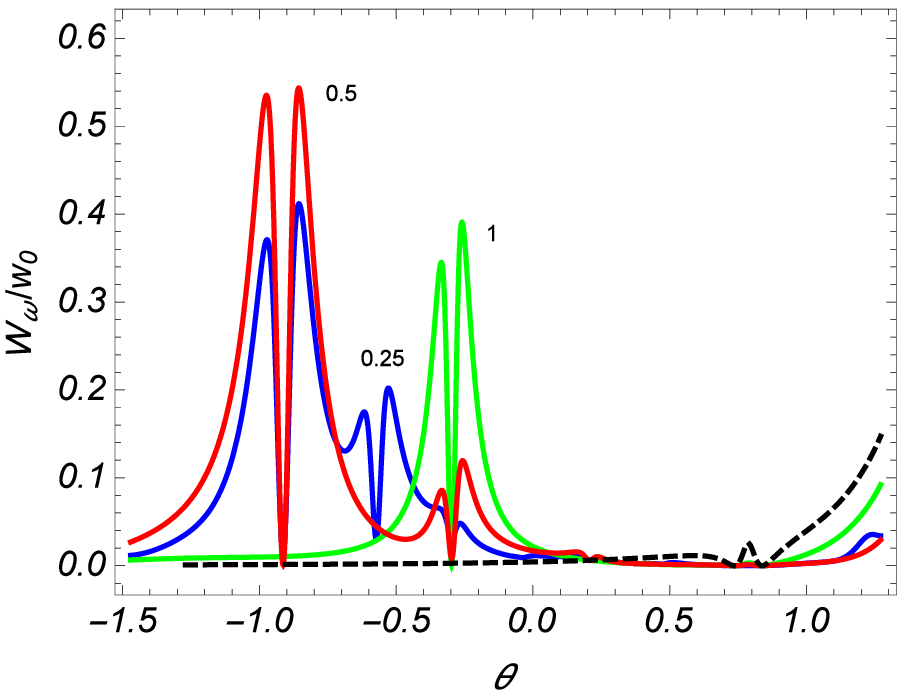,width=7.cm,height=5.5cm} & \quad %
\epsfig{figure=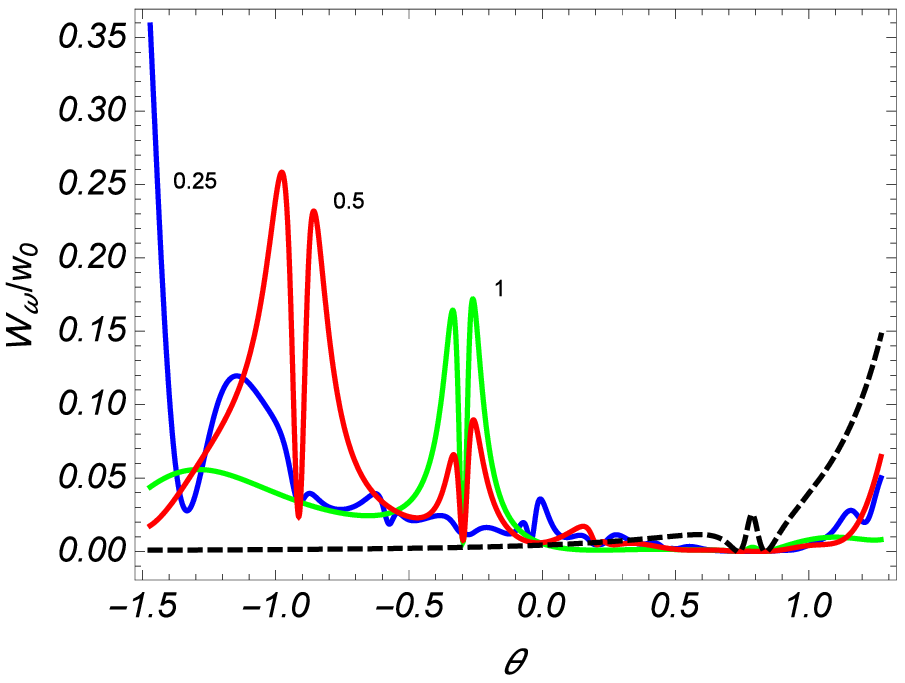,width=7.cm,height=5.5cm}%
\end{tabular}%
\end{center}
\caption{The spectral-angular density of the radiated energy, as a function
of the radiation angle, for $\protect\beta _{0}\approx 0.998694$ (the
electron energy 10 MeV in the case $\protect\varepsilon _{0}=1$) and for
separate values of the ratio $\protect\lambda /L$ (numbers near the curves).
Figure \protect\ref{fig3}a corresponds to the profile function (\protect\ref%
{x0Grat}) with $b=L/2$ and Fig. \protect\ref{fig3}b corresponds to the
function (\protect\ref{x0wave}) with $\protect\omega _{s}=0$. The dashed
curves correspond to the radiation on a flat boundary. The graphs are
plotted for $\protect\theta _{g}=\protect\pi /4$ and $a/L=0.5$. }
\label{fig3}
\end{figure}

In figures below we plot the ratio $W_{\omega }/w_{0}$, with $w_{0}=\alpha
\hbar \left( \Delta \varepsilon /\varepsilon _{0}\right) ^{2}/\sqrt{%
\varepsilon _{0}}$ and $\alpha =q^{2}/(\hbar c)$, as a function of the
projection angle $\theta $ between the $x$-axis and the wave vector of the
radiation: $k_{x}=k\cos \theta $, $k_{z}=k\sin \theta $. For the radiation
propagating in the medium with permittivity $\varepsilon _{0}$ one has $-\pi
/2<\theta <\pi /2$. Recall that the validity of the approximation we used is
constrained by the conditions (\ref{CondAd}). We can express these
conditions in terms of the angles $\theta $ and $\theta _{g}$ by taking into
account that $\theta _{v}=\pi -\theta _{g}$ and $\cos \theta _{0}=-\cos
(\theta +\theta _{g})$. In particular, the approximation fails for the
values of $|\theta |$ close to $\pi /2$. Note that, for a static interface,
the ratio $W_{\omega }/w_{0}$ depends on the velocity of the charge, $v$,
and on the permittivity $\varepsilon _{0}$ through the parameter $\beta _{0}$%
. In this case the condition for the angular locations of the peaks, for a
given value of $\lambda /L$, is obtained from (\ref{CondPeak2}) and, up to
the terms of the order $1/\gamma _{0}^{2}$, is written in the form%
\begin{equation}
\sin \theta \approx \sin (\theta _{g})/\beta _{0}+l\lambda /L.
\label{CondPeak3}
\end{equation}%
In figure \ref{fig3}, for the value $\beta _{0}\approx 0.998694$, we have
plotted the quantity $W_{\omega }/w_{0}$ as a function of the angle $\theta $
for different values of the ratio $\lambda /L$ (numbers near the curves).
For $\varepsilon _{0}=1$ this value of $\beta _{0}$ corresponds to the
electron energy $E_{e}=10$ MeV. Figure \ref{fig3}a corresponds to the
profile function (\ref{x0Grat}) with $b=L/2$ and Fig. \ref{fig3}b
corresponds to the function (\ref{x0wave}) with $\omega _{s}=0$. The dashed
curves correspond to the radiation on a flat boundary ($a=0$). For the
incidence angle with respect to the normal and for the ratio $a/L$ we have
taken $\theta _{g}=\pi /4$ and $a/L=0.5$. It can be checked that the angular
locations of the peaks for the graphs presented in figure \ref{fig3} are
well described by the formula (\ref{CondPeak3}). For example, in the case of
the radiation on the triangular grating at the wavelength corresponding to $%
\lambda /L=0.5$ the angular density of the radiation intensity is maximum
around the angle $\theta \approx -0.91$ which is obtained from (\ref%
{CondPeak3}) for the order of diffraction $l=-3$. For the radiation with $%
\lambda /L=0.5$ the maximum is around $\theta \approx -0.3$ and corresponds
to the order of diffraction $l=-1$. Note that for the latter case there are
no peaks for $l\neq -1$.

\begin{figure}[tbph]
\begin{center}
\begin{tabular}{cc}
\epsfig{figure=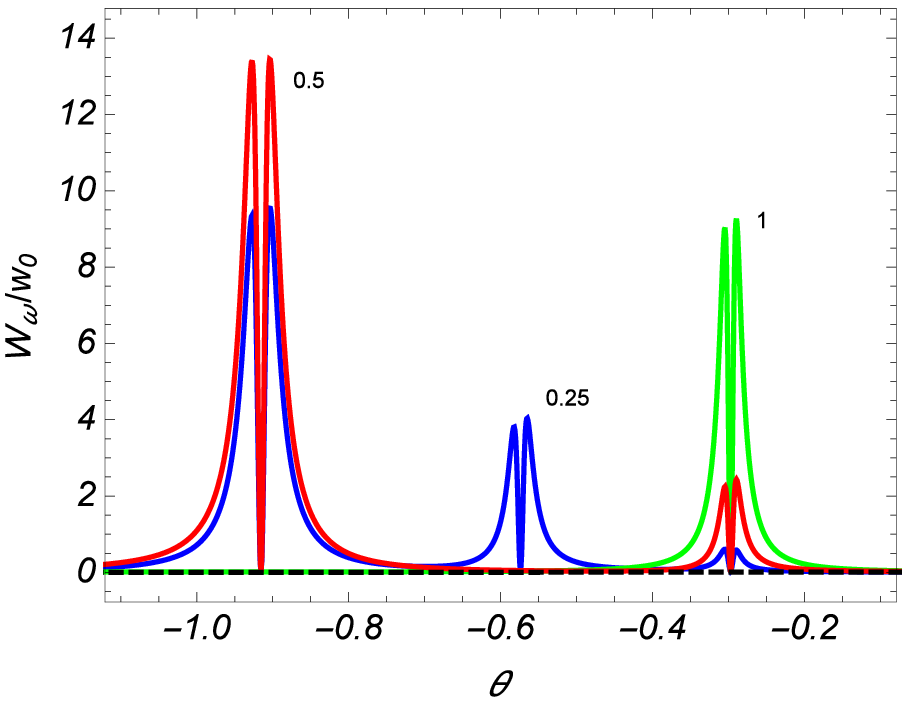,width=7.cm,height=5.5cm} & \quad %
\epsfig{figure=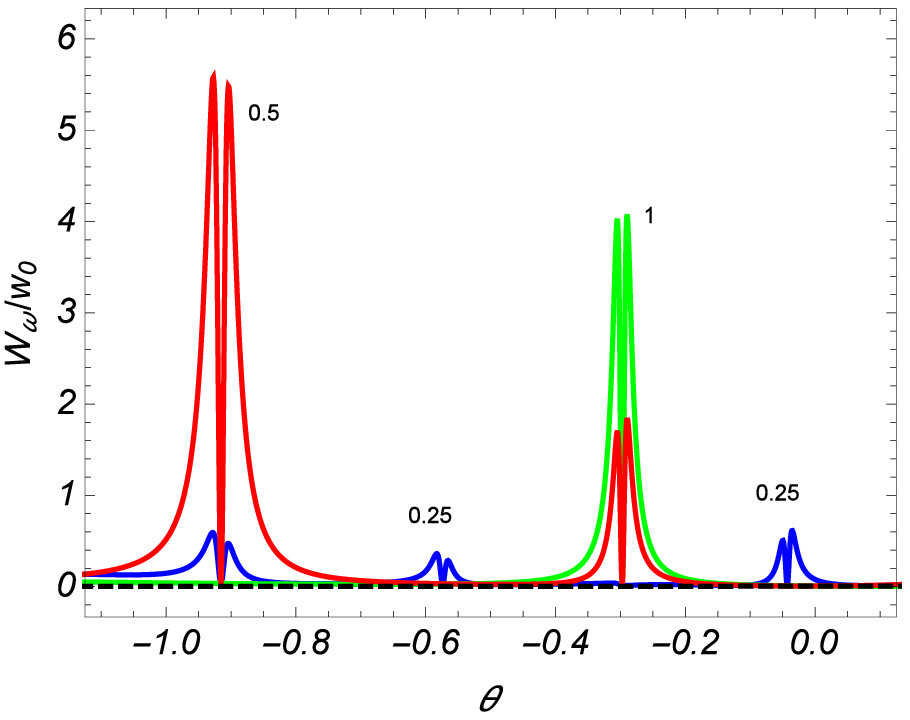,width=7.cm,height=5.5cm}%
\end{tabular}%
\end{center}
\caption{The same as in figure \protect\ref{fig3} for $\protect\beta %
_{0}\approx 0.999948$ (the electron energy 50 MeV in the case $\protect%
\varepsilon _{0}=1$). The numbers near the curves correspond to the values
of $\protect\lambda /L$.}
\label{fig4}
\end{figure}

Figure \ref{fig4} presents the same graphs as in figure \ref{fig3} for $%
\beta _{0}\approx 0.999948$. In the case $\varepsilon _{0}=1$ the latter
corresponds to the electron energy $E_{e}=50$ MeV. As is seen, in accordance
with the analytic estimate given above, with the increase of the energy the
heights of the peaks increase by the factor $\gamma _{0}^{2}$, whereas the
widths decrease by the factor $\gamma _{0}$. For $\gamma _{0}^{2}\gg 1$, the
angles around which the peaks are located are not sensitive to the value of $%
\beta _{0}$ and they are approximately the same as those for \ref{fig3}.

\begin{figure}[tbph]
\begin{center}
\begin{tabular}{cc}
\epsfig{figure=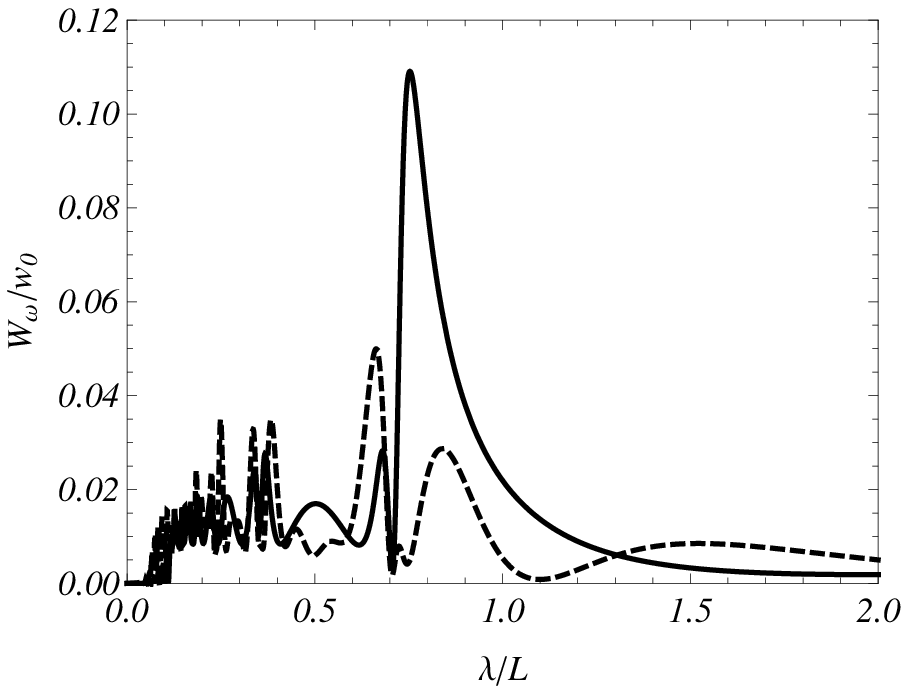,width=7.cm,height=5.5cm} & \quad %
\epsfig{figure=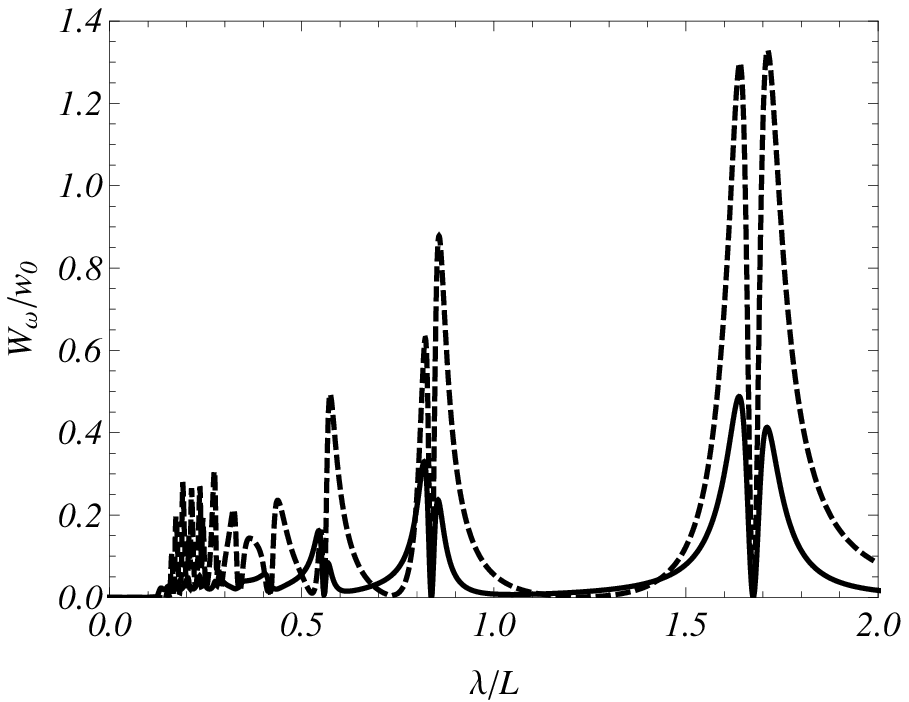,width=7.cm,height=5.5cm}%
\end{tabular}%
\end{center}
\caption{The dependence of the spectral-angular density of the radiated
energy on the ratio $\protect\lambda /L$ for the radiation angles $\protect%
\theta =0$ (Fig. \protect\ref{fig5}a) and $\protect\theta =-5\protect\pi /12$
(Fig. \protect\ref{fig5}b). The graphs are plotted for the function (\protect
\ref{x0Grat}) with $b=L/2$ (full curves) and for the function (\protect\ref%
{x0wave}) with $\protect\omega _{s}=0$ (dashed curves). The values of the
other parameters are the same as in figure \protect\ref{fig3}.}
\label{fig5}
\end{figure}

Figure \ref{fig5} displays the dependence of the spectral-angular density of
the radiated energy on the ratio $\lambda /L$ for the function (\ref{x0Grat}%
) with $b=L/2$ (full curves) and for the function (\ref{x0wave}) with $%
\omega _{s}=0$ (dashed curves). The graphs in Fig. \ref{fig5}a and \ref{fig5}%
b are plotted for the radiation angles $\theta =0$ and $\theta =-5\pi /12$,
respectively. For both cases $\beta _{0}\approx 0.998694$ (the electron
energy $E_{e}=10$ MeV in the case $\varepsilon _{0}=1$), $\theta _{g}=\pi /4$
and $a/L=0.5$. Again, we can see that the values of the radiation
wavelengths around of which the peaks are located are well described by the
formula (\ref{CondPeak3}). In particular, for the peak with the maximal
value of the wavelength one has $\lambda \approx L|\sin \theta -\sin (\theta
_{g})/\beta _{0}|$.

\begin{figure}[tbph]
\begin{center}
\begin{tabular}{cc}
\epsfig{figure=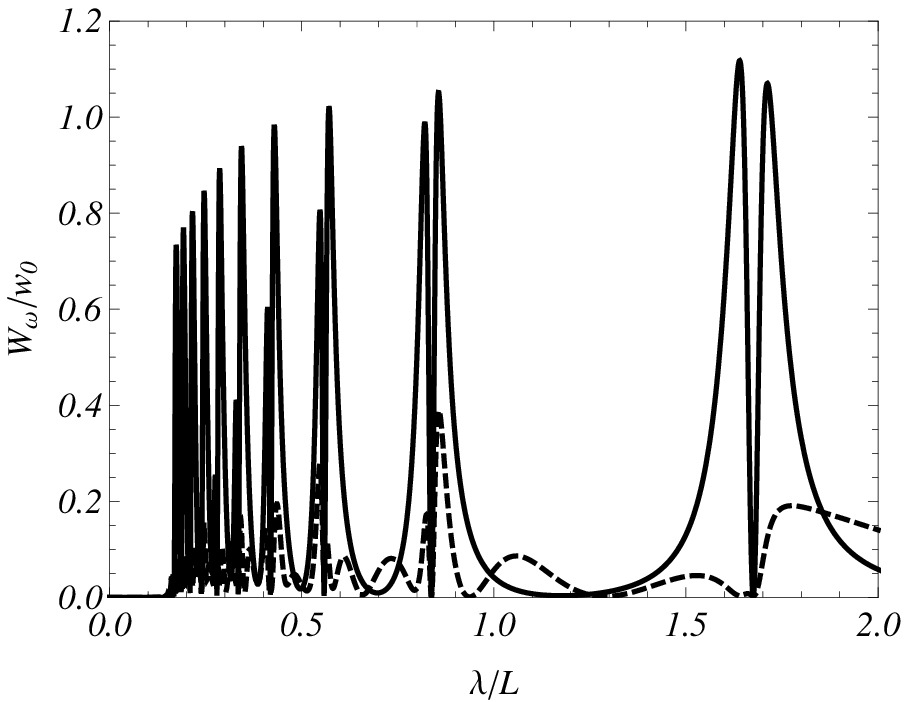,width=7.cm,height=5.5cm} & \quad %
\epsfig{figure=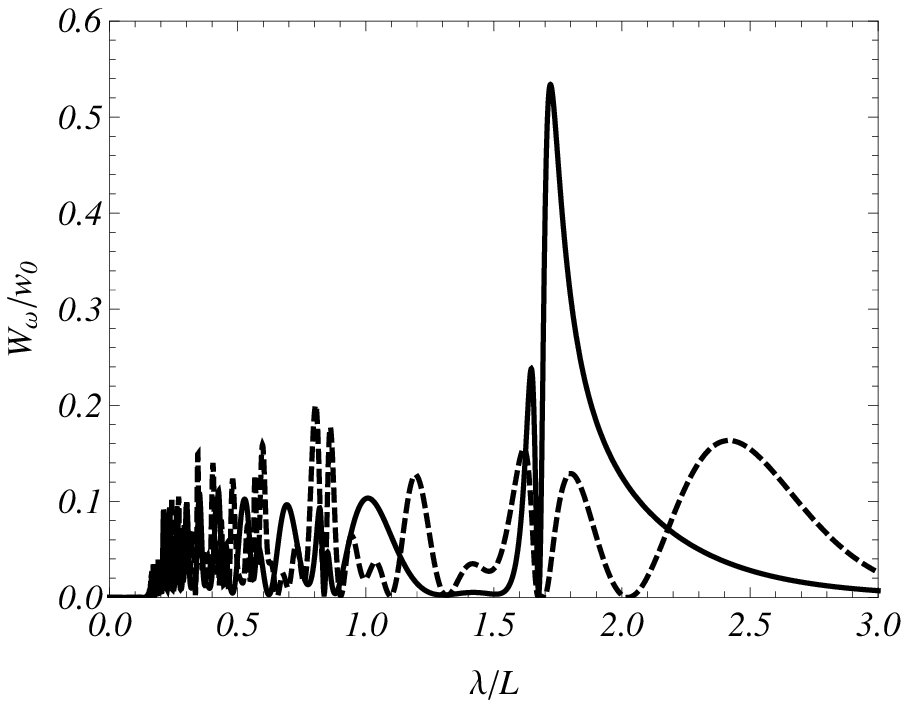,width=7.cm,height=5.5cm}%
\end{tabular}%
\end{center}
\caption{The same as in Fig. \protect\ref{fig5}b for $a/L=1$ (Fig. \protect
\ref{fig6}a) and $a/L=2$ (Fig. \protect\ref{fig6}b).}
\label{fig6}
\end{figure}

The same graphs, as in the right panel of figure \ref{fig5}, are plotted in
Fig. \ref{fig6} for the values $a/L=1$ (Fig. \ref{fig6}a) and $a/L=2$ (Fig. %
\ref{fig6}b). As is seen from the figures, the difference of the wavelengths
for the neighboring peaks decreases with decreasing wavelength. This is in
agreement of the general estimate (\ref{CondPeak3}) for the spectral-angular
locations of the peaks. The right peaks correspond to first order of
diffraction $l=-1$.

\begin{figure}[tbph]
\begin{center}
\epsfig{figure=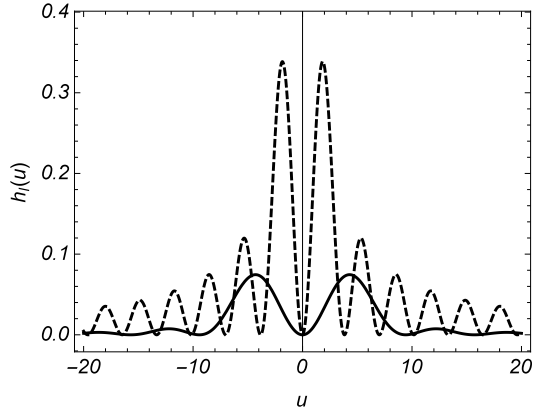,width=7.cm,height=5.5cm}
\end{center}
\caption{The function $h_{l}(u)$ for $l=-1$ in the expression for the
spectral-angular density of the radiated energy in the cases of the
triangular (with $b=L/2$, full curve) and sinusoidal (dashed curve)
profiles. }
\label{fig7}
\end{figure}

The radiation intensity at the peaks is mainly determined by the factors $%
|f_{l}(\omega K/v)|^{2}$ and $U^{-2}$ in (\ref{Wang}). As is seen from (\ref%
{fnKx}), the function $f_{l}(K_{x})$ depends on the amplitude $a$ of the
function $x_{0}(z)$ in the form of the dimensionless combination $aK_{x}$.
Hence, the dependence on the amplitude enters in the expression for the
radiation intensity through the function $h_{l}(a\omega K/v)=|f_{l}(\omega
K/v)|^{2}$. By taking into account the expression (\ref{K}), for the special
case under consideration corresponding to $\phi _{v}=0$, the argument of the
function $h_{l}(u)$ in the expression for the radiation intensity is
presented as%
\begin{equation}
aK\frac{\omega }{v}=-\frac{2\pi a}{L\cos \theta _{g}}\left\{ \frac{L}{%
\lambda }[1/\beta _{0}+\cos (\theta +\theta _{g})]+l\sin \theta _{g}\right\}
.  \label{hlarg}
\end{equation}%
For the profiles we have discussed in numerical examples, the expressions
for the function $h_{l}(u)$ directly follow from (\ref{fnGrat}) and (\ref%
{flSurfWave}). In figure \ref{fig7}, the functions $h_{l}(u)$ are plotted
for the profiles (\ref{x0Grat}) with $b=L/2$ (full curve) and (\ref{x0wave})
(dashed curve) in the case $l=-1$. The graphs for higher orders of
diffraction have a similar structure.
The values for $|u|$, at which the function $h_{l}(u)$ takes its first
maximum (with respect to $u=0$), increases with increasing $|l|$, whereas
the value of the function at the maximum decreases. Let $u=u_{l}$ is the
maximum point of the function $h_{l}(u)$ for a given $l$. Now, from the
analysis given above, it follows that the maximal radiation intensity at the
peaks, determined by the condition (\ref{CondPeak2}), is obtained for the
values of the amplitude given by $a=vu_{l}/(\omega K)$. An explicit
expression in terms of the angles $\theta $ and $\theta _{g}$ is obtained by
taking into account (\ref{hlarg}). Note that, at the peaks, we can exclude
the ratio $\lambda /L$ from the corresponding expression by using the
relation (\ref{CondPeak2}). In the case of the profile function (\ref{x0wave}%
), $u_{l}$ coincides with the first zero of the function $J_{l}^{\prime }(u)$%
. In particular, for $l=-1$ one has $u_{l}\approx 1.84$. With this value of $%
u_{l}$ and for $\theta _{g}=\pi /4$, $\theta =-5\pi /12$, $\beta _{0}\approx
0.998694$, the maximum radiation at the peaks is obtained for $a\approx 0.5$%
. This case corresponds to the dashed curve in Fig. \ref{fig5}b. For the
triangular profile with $b=L/2$, in the case of diffraction order $l=-1$ one
has $u_{l}\approx 4.3$ and for the same values of the parameters $\theta
_{g} $, $\theta $, $\beta _{0}$, the maximum radiation at the corresponding
peak is realized for $a\approx 1.2$. The corresponding graph for the
spectral density is similar to the full curve in Fig. \ref{fig6}a.

\section{Conclusion}

\label{sec:Conc}

We have considered transition radiation of a charged particle on an
interface of two dielectric media having an arbitrary non-stationary
profile. The exact solution of this problem is complicated and we have
employed an approximate scheme which assumes that the dielectric
permittivities of the media are close. With this assumption, the
spectral-angular densities of the radiated energy on separate polarizations
are given by Eq. (\ref{Wp3}) and the total radiated energy is determined by
Eq. (\ref{W2}). The dependence on the geometry of the separating boundary
enters in these expressions through the functions $f_{mnl}(K_{x})$ which are
the coefficients of the Fourier expansion of the function $%
e^{iK_{x}\,x_{0}(y,z,t)}$. For high-energy particles strong peaks may appear
in the spectral-angular density of the radiated energy. The locations of the
peaks are determined by the condition (\ref{CondPeak}). This condition is
completely determined by the periodicity properties of the interface and
does not depend on the specific form of the profile function.

The general expressions for the radiated energy are simplified in a special
case of the profile function given by (\ref{x0Run}). In this case the triple
summation is reduced to the single one and the corresponding formulas are
obtained by the replacements (\ref{ReplSp}). A physical realization of this
sort of profile function could be a surface wave excited on the boundary
between two media. In terms of the spherical angular coordinates for the
vectors $\mathbf{k}$ and $\mathbf{v}$, the expressions for the radiated
energy are presented in the form (\ref{Wpang}) and (\ref{Wang}). As
examples, in the numerical evaluation we have considered two special cases
of the profile function $x_{0}(u)$: triangular grating with the parameters
displayed in figure \ref{fig1} and sinusoidal profile. Keeping in mind
applications for beam diagnostics, in the numerical evaluation we have
investigated the backward radiation for these two examples. Instead of a
single peak in the backward transition radiation on a flat interface, for
periodic interface one has a set of peaks. The number and the locations of
the peaks depend on the incidence angle of the charge and on the period of
the interface. The heights of the peaks in the spectral-angular distribution
increase with increasing energy of the radiating particle as $\gamma
_{0}^{2} $, whereas the widths decrease as $1/\gamma _{0}$. The results
above show that the periodic structure on the interface may serve as an
additional tool for the control of spectral-angular distribution of the
backward transition radiation. The corresponding features may be useful in
beam and surface diagnostics.

\end{document}